\documentclass[10pt,twocolumn,letterpaper]{article}

\usepackage[pagenumbers,applications]{wacv}      

 \usepackage{comment}
\definecolor{wacvblue}{rgb}{0.21,0.49,0.74}
\usepackage[pagebackref,breaklinks,colorlinks,allcolors=wacvblue]{hyperref}
\usepackage{xcolor}
\usepackage{booktabs}
\usepackage{multirow}
\usepackage{graphicx}
\usepackage{amssymb}
\usepackage{booktabs}
\usepackage{graphicx}
\usepackage{amsmath}
\usepackage{amssymb}
\def\wacvPaperID{1254} 
\def\confName{WACV}
\def\confYear{2027}

\title{SciFusion: Physics-Guided Compositional Diffusion for Multimodal 3D Scientific Image Generation}

\author{
Luoxi Wang$^{1,2}$ \quad
Alessio Quercia$^2$ \quad
Hanno Scharr$^{2,\dagger}$ \qquad
Arya Bangun$^{2,\dagger}$ \\[1ex]
$^1$Princeton University, Princeton, NJ, USA \\
$^2$Forschungszentrum J\"ulich, J\"ulich, Germany \\[4pt]
\shortstack{
{\tt\small luoxi.wang@princeton.edu} \\
{\tt\small \{a.quercia,h.scharr,a.bangun\}@fz-juelich.de}
}
}

\begin{document}
\maketitle

\begingroup
\renewcommand{\thefootnote}{\ensuremath{\dagger}}
\footnotetext{These authors contributed equally as senior authors.}
\endgroup
\begin{abstract}
Existing 3D generative approaches are typically developed for individual imaging modalities, requiring substantial modality-specific data and limiting their ability to exploit shared structure across scientific imaging domains. We present SciFusion, a multimodal framework for unified 3D scientific image generation built on a latent diffusion backbone, with two key contributions. First, we introduce a compositional conditioning scheme that factorizes each modality into domain, object, and contrast representations. Sharing these factors across related modalities enables knowledge transfer
across heterogeneous imaging domains and allows the factors to be recombined
at inference time to synthesize unseen domain–object–contrast compositions
in a zero-shot manner. Second, we incorporate domain-specific physics-informed objectives—Fourier spectral consistency for MRI, Radon projection consistency for CT, and multislice diffraction consistency for electron microscopy—to encourage generated samples to respect modality-dependent image formation processes. Extensive experiments across 15 medical and scientific imaging modalities demonstrate that SciFusion improves generation fidelity and cross-domain generalization over modality-specific and flattened multimodal conditioning strategies, while enabling zero-shot synthesis of unseen compositions without training on those target compositions. These results demonstrate that compositional, physics-guided conditioning provides a scalable framework for unified 3D scientific image generation, particularly for emerging modalities where large-scale training data are difficult to obtain.
\noindent{Project page:}
\url{https://ixoul.github.io/SciFusion-Demo/}
\end{abstract}    

\section{Introduction}
\label{sec:intro}
Three-dimensional scientific imaging plays a critical role in biomedical
research and scientific discovery, covering diverse modalities such as
magnetic resonance imaging (MRI), computed tomography (CT), and electron
microscopy (EM). However, collecting large-scale 3D imaging datasets remains
challenging due to expensive acquisition procedures, specialized imaging
equipment, and privacy constraints in medical applications. These limitations
result in fragmented datasets across different imaging domains, restricting
the development of data-driven approaches.

Generative models provide a promising solution by learning complex data
distributions and synthesizing realistic samples to alleviate data scarcity,
support data augmentation, and enable privacy-preserving data
sharing~\cite{tudosiu2024realistic, khader2023denoising}. Recent diffusion
models have achieved remarkable success in image generation by learning to
reverse a gradual noising process~\cite{ddpm2020, song2020score}. Latent
diffusion further reduces the computational and memory requirements of 3D
generation by operating in a compressed latent
space~\cite{ldm2022, blattmann2023alignlatentshighresolutionvideo}, enabling
more efficient synthesis of complex 3D data. Despite these advances, most
existing 3D generative models are designed for individual datasets, objects,
or imaging modalities. Such modality-specific models require separate
training processes and large amounts of dedicated data, limiting knowledge
transfer across related imaging domains. Scientific imaging, however, is
inherently heterogeneous: MRI modalities differ in anatomical targets and
tissue contrasts~\cite{Chen2024TowardsGT}, CT captures different structures
through X-ray attenuation, and EM reveals microscopic structures with
distinct frequency characteristics. A unified generative model for
heterogeneous scientific imaging therefore needs to learn shared
representations across modalities while preserving domain-specific imaging
properties.

A key challenge toward such a unified model is how to effectively represent
and combine heterogeneous imaging information. Treating each modality as an
independent class limits generalization to unseen conditions, while directly
merging all modalities may introduce ambiguous conditioning and weaken
modality-specific characteristics. A more expressive representation should
capture the underlying factors that define an imaging modality and allow
these factors to be recombined for flexible generation.

In this work, we present SciFusion, a multimodal 3D scientific image
generation framework built upon a latent diffusion backbone that combines
physics-guided objectives with compositional conditioning across MRI, CT,
and EM modalities. Our contributions can be summarized as follows:
\begin{itemize}
    \item We develop a unified framework that learns a shared latent
    representation using a vector-quantized autoencoder and performs
    volumetric generation with a conditional latent diffusion model.
    \item We introduce a compositional conditioning scheme that factorizes
    modality information into domain, object, and contrast embeddings,
    enabling zero-shot generation of unseen domain, object, and contrast
    combinations without additional modality-specific training data.
    \item We design domain-specific physical guidance, including power-spectrum
consistency for MRI, Radon-domain constraints for CT, and diffraction-pattern
constraints for EM, to preserve imaging characteristics.
    \item We conduct comprehensive evaluations across observed and unseen
    modality combinations, demonstrating transferable generation and improved
    consistency with domain-specific physical properties.
\end{itemize}
\section{Related Works}
\label{sec:relatedworks}
\subsection{3D Volumetric Image Generation}
Diffusion models have recently enabled high-quality synthesis of
three-dimensional medical and scientific images. Medical diffusion performs
volumetric generation directly in image space and demonstrates applications
to MRI and CT synthesis~\cite{khader2023denoising}. To improve scalability,
recent methods introduce latent representations for efficient volumetric
generation. 3D MedDiffusion and MedLoRD employ pretrained autoencoders to
support latent diffusion of 3D medical volumes~\cite{3dmeddiff,medlord}.
MAISI enables high-resolution CT generation with flexible compression and
anatomical conditioning~\cite{maisi-v1}. MAISI-v2 further improves sampling
efficiency and conditioning accuracy using rectified flow and contrastive
learning~\cite{maisi-v2}. Despite these advances, existing methods mainly
target specific medical domains or predefined modality conditions, limiting
their ability to learn transferable representations across heterogeneous
scientific imaging modalities.

\subsection{Physics-Guided Generative Models}
A complementary line of work incorporates physical priors to
improve the fidelity and physical plausibility of generated scientific
images, benefiting tasks such as reconstruction and cross-modality
generation~\cite{chung2022score, song2021scorebased, Bangun_2025}. For
multimodal MRI synthesis, a physics-informed latent diffusion framework first
generates latent tissue property maps and subsequently applies a physical
signal model to synthesize different MRI contrasts, including unseen
modalities~\cite{physicsinformedldm}, demonstrating the potential of
combining generative models with acquisition physics for cross-modality
synthesis. Beyond medical imaging, physics-guided diffusion priors have also
been incorporated for multi-slice reconstruction in scientific imaging,
showing the benefit of physical constraints for recovering structurally
consistent images~\cite{valdy2026physics}. However, these physics-guided
generative approaches are primarily developed within individual imaging
domains, where physical constraints are closely coupled with a specific
acquisition process. Extending physics-guided generation to heterogeneous
scientific imaging remains challenging, as different modalities follow
distinct physical formation mechanisms, motivating a unified framework that
combines transferable multimodal representations with domain-specific
physical guidance.
\begin{figure*}[t]
  \centering
   \includegraphics[width=0.9\linewidth]{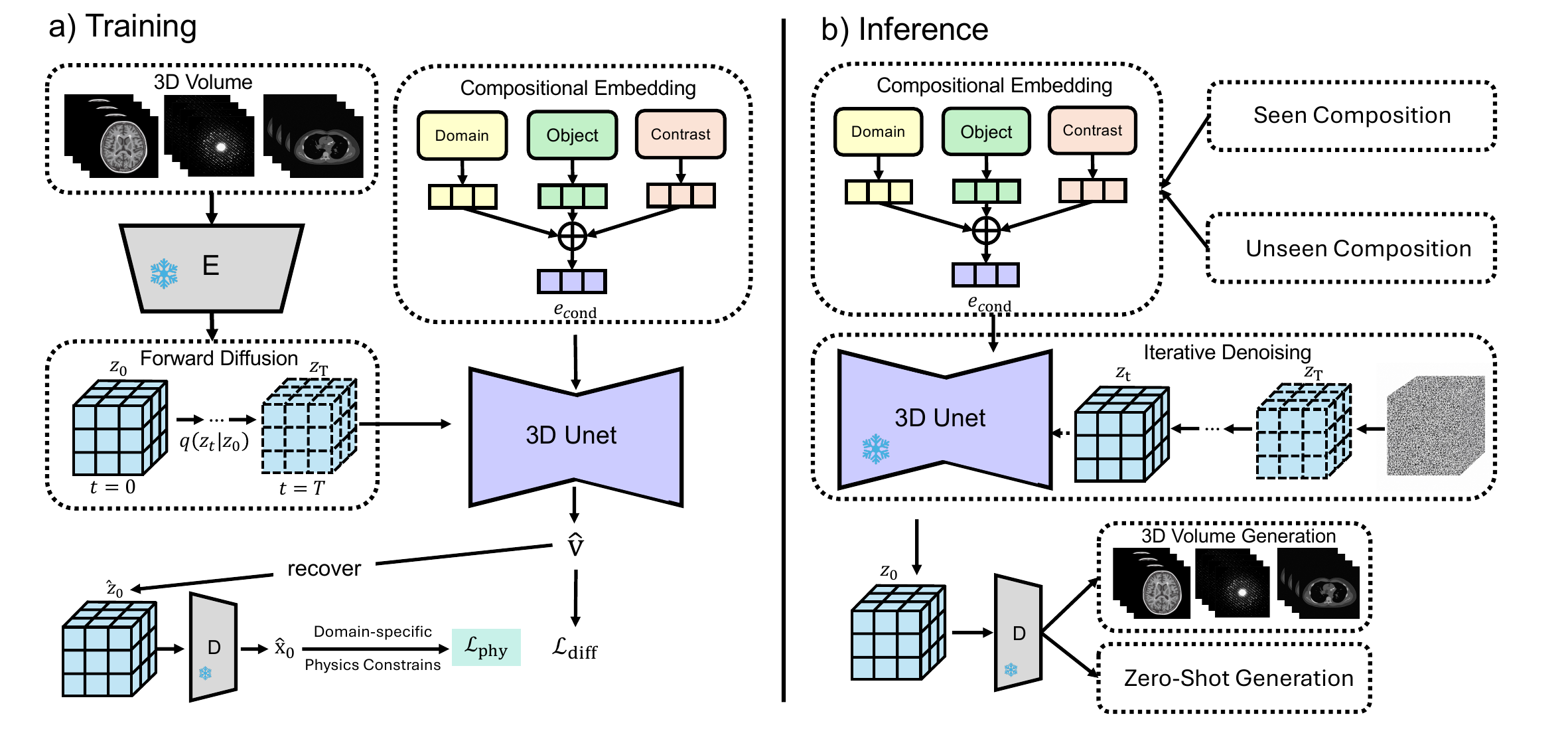}
    \caption{
Overview of SciFusion.
(a) Two-stage training with shared latent representation learning, compositional conditioning, and domain-specific physics guidance.
(b) Inference via recombination of learned conditioning components, enabling both seen-modality generation and zero-shot synthesis of unseen modality combinations.
}
   \label{fig:architecture}
\end{figure*}
\section{Method}
SciFusion builds upon the latent diffusion framework~\cite{ldm2022}
and the MedLoRD architecture~\cite{medlord} for efficient high-resolution
3D medical image synthesis (Fig.~\ref{fig:architecture}). Training proceeds
in two stages: a VQGAN is first pretrained on the integrated multi-modal
dataset without modality conditioning to learn a shared latent
representation across heterogeneous imaging domains (Sec.~\ref{sec:latent_diffusion}); a conditional 3D
latent diffusion model is then trained on the frozen VQGAN latents, with
modality information injected through compositional conditioning
(Sec.~\ref{sec:composition}) and domain-specific physical guidance
(Sec.~\ref{sec:physics}). At inference, the learned compositional
representation supports both seen-composition and zero-shot generation of
unseen domain--object--contrast combinations. Detailed training configurations and exact latent dimensions are provided
in Sec.~\ref{supp:training} of the supplementary material.
\subsection{Latent Diffusion Framework}
\label{sec:latent_diffusion}
Each volumetric sample is represented as $\mathbf{x}\in\mathbb{R}^{C\times S\times N \times N}$,
where the dimensions correspond to the channel, depth, height, and width,
respectively. The pretrained, frozen VQGAN serves as the first-stage
compression model: its encoder $\mathcal{E}$ maps the input volume to a
lower-dimensional latent representation, $\mathbf{z}_0=\mathcal{E}(\mathbf{x})$,
where $\mathbf{z}_0\in\mathbb{R}^{C\times S \times L \times L}$ and $L \ll N$.
Only the spatial dimensions are compressed, while the depth dimension $S$ is
preserved to retain volumetric continuity across slices. The corresponding
decoder reconstructs the volumetric image as $\hat{\mathbf{x}}=\mathcal{D}(\mathbf{z}_0)$.
During diffusion training, a timestep $t$ is randomly sampled, and Gaussian
noise $\boldsymbol{\epsilon}\sim\mathcal{N}(\mathbf{0},\mathbf{I})$ is added to
the clean latent representation according to a noise schedule:
\begin{equation}
\mathbf{z}_t
=
\alpha_t \mathbf{z}_0
+
\sigma_t \boldsymbol{\epsilon},
\end{equation}
where $\alpha_t$ and $\sigma_t$ control the contributions of the clean latent
signal and Gaussian noise at timestep $t$. Following the $v$-prediction
formulation, the corresponding training target is
\begin{equation}
\mathbf{v}_t
=
\alpha_t \boldsymbol{\epsilon}
-
\sigma_t \mathbf{z}_0.
\end{equation}
A conditional 3D U-Net $f_\theta$ predicts the velocity target from the noisy
latent representation, diffusion timestep, and conditioning representation:
\begin{equation}
\hat{\mathbf{v}}_\theta
=
f_\theta
\left(
\mathbf{z}_t,
t,
\mathbf{e}_{\mathrm{cond}}
\right),
\end{equation}
where $\mathbf{e}_{\mathrm{cond}}$ denotes the compositional conditioning
embedding introduced in the following subsection. The latent diffusion objective
is then defined as
\begin{equation}
\mathcal{L}_{\mathrm{diff}}
=
\mathbb{E}_{\mathbf{z}_0,t,\boldsymbol{\epsilon}}
\left[
\left\|
\mathbf{v}_t
-
\hat{\mathbf{v}}_\theta
\right\|_2^2
\right].
\end{equation}
Unlike the original single-domain formulation of MedLoRD, our framework
employs a shared diffusion backbone to jointly model heterogeneous imaging
modalities, while modality-specific generation is controlled through the
proposed compositional conditioning mechanism.
\subsection{Compositional Conditioning Mechanism}
\label{sec:composition}
A conventional multimodal diffusion model typically assigns an independent
learnable embedding to each imaging modality. Although this flattened
conditioning strategy can distinguish modalities observed during
training~\cite{3dmeddiff}, it treats each modality as an unrelated category,
without exploiting shared structure across heterogeneous scientific images.
To address this limitation, we represent each modality as a composition of
three semantic factors: imaging domain, anatomical or structural object, and
imaging contrast. Specifically, the condition of a volumetric sample is
represented as $\mathbf{c}=(c_{\mathrm{dom}},c_{\mathrm{obj}},c_{\mathrm{con}})$,
where $c_{\mathrm{dom}}$, $c_{\mathrm{obj}}$, and $c_{\mathrm{con}}$ denote the
imaging domain, target object, and acquisition contrast, respectively.
We introduce three independent learnable embedding tables $\mathcal{E}_{\mathrm{dom}}$,
$\mathcal{E}_{\mathrm{obj}}$, $\mathcal{E}_{\mathrm{con}}$, yielding embeddings
$\mathbf{e}_{k}=\mathcal{E}_{k}(c_{k})\in\mathbb{R}^{d_{\mathrm{cond}}}$ for
$k\in\{\mathrm{dom},\mathrm{obj},\mathrm{con}\}$, where $d_{\mathrm{cond}}$ is
the conditioning dimension.
The final conditioning representation is obtained through element-wise addition:
\begin{equation}
\mathbf{e}_{\mathrm{cond}}
=
\mathbf{e}_{\mathrm{dom}}
+
\mathbf{e}_{\mathrm{obj}}
+
\mathbf{e}_{\mathrm{con}}.
\end{equation}
The resulting vector is treated as a single conditioning token,
$\mathbf{e}_{\mathrm{cond}}\in\mathbb{R}^{1\times d_{\mathrm{cond}}}$, and is
injected into the conditional 3D U-Net through its cross-attention layers.
For a batch of $B$ volumetric samples, the complete conditioning tensor has
the shape $\mathbf{E}_{\mathrm{cond}}\in\mathbb{R}^{B\times1\times d_{\mathrm{cond}}}$.
Compared with a flattened modality embedding, this factorized representation
allows related modalities to share parameters at multiple semantic levels.
Modalities from the same imaging domain reuse the domain embedding, while
modalities depicting the same object or using the same contrast share the
corresponding object or contrast embedding. This encourages the model to
learn both modality-specific characteristics and common structures across
heterogeneous imaging modalities. Additionally, it allows unseen modality combinations
to be constructed directly from factors observed during training, without
introducing an additional modality-specific embedding. This property provides
the basis for the compositional zero-shot generation evaluated in our
experiments.
\subsection{Physics-Informed Training Objective}
\label{sec:physics}
While the latent denoising objective encourages accurate recovery of the
clean latent representation, it does not explicitly enforce the physical
acquisition constraints of different imaging domains. Given the predicted
noise, we first recover the clean latent estimate $\hat{\mathbf{z}}_0$ and
decode it into a predicted volume $\hat{\mathbf{x}}_0$. For each sample, a
domain-dependent differentiable operator $\mathcal{P}_d$ is applied to both
the reference and predicted volumes, with logarithmic scaling applied to
reduce their dynamic range:
\begin{equation}
\mathcal{L}_{\mathrm{phys}}
=
\frac{1}{B}
\sum_{b=1}^{B}
\left\|
\mathcal{P}_{d_b}(\mathbf{x}_b)
-
\mathcal{P}_{d_b}(\hat{\mathbf{x}}_{0,b})
\right\|_1,
\end{equation}
where $d_b$ denotes the imaging domain of the $b$-th sample, with
$\mathcal{P}_d$ instantiated as one of $\mathcal{P}_{\mathrm{MRI}}$,
$\mathcal{P}_{\mathrm{CT}}$, or $\mathcal{P}_{\mathrm{EM}}$, as described below This
formulation allows different physical or spectral constraints to be applied
within the same mixed-domain training batch.
\paragraph{Magnetic Resonance Imaging (MRI).}
For MRI, since raw data is acquired in k-space, we constrain the slice-wise
spatial-frequency distribution using a logarithmic Fourier power-spectrum
representation:
\begin{equation}
\mathcal{P}_{\mathrm{MRI}}(\mathbf{x}) =
\log_{10}\!\left(
1+
\left\lvert\!
\mathcal{F}_{2\mathrm{D}}(\mathbf{x})
\right\rvert^2
\right).
\label{eq:mri_spectral_operator}
\end{equation}
This spectral loss aligns the frequency content of generated and reference
MRI volumes.
\paragraph{Computed Tomography (CT).}
For CT, we impose consistency in the tomographic projection domain using a
differentiable Radon transform over multiple projection angles to obtain a
sinogram representation~\cite{HeRadonInversion,xu2025DeepRadon}:
\begin{equation}
\mathcal{P}_{\mathrm{CT}}(\mathbf{x})
=
\log_{10}\!\left(1+\left\lvert\mathcal{R}(\mathbf{x})\right\rvert^2\right),
\label{eq:ct_projection_operator}
\end{equation}
where $\mathcal{R}$ denotes the differentiable Radon transform. This loss
enforces consistency between the projection measurements of generated and
reference CT volumes.
\paragraph{Electron Microscopy (EM).}
For EM, we apply a diffraction-inspired differentiable multislice
operator~\cite{cowley1957scattering, cowley1959scattering, cowley1959scattering2, ishizuka1977new, kirkland1998advanced}
using plane wave illumination $\psi_0$, which is sequentially modulated by
the slices and propagated using a Fresnel kernel. The wave function after the
$n$-th slice is computed as
\begin{equation}
\psi_n = \mathcal{F}^{-1}_{2D}\!\left[
\mathcal{F}_{2D}\!\left(\psi_{n-1}\exp(iV_n)\right)P
\right].
\label{eq:multislice_propagation}
\end{equation}
Here, $V_n$ denotes the $n$-th slice of volumetric data (i.e., the projected
electrostatic potential), and $\mathcal{F}_{2D}$ and $\mathcal{F}^{-1}_{2D}$
denote the two-dimensional Fourier and inverse Fourier transforms,
respectively, while $P$ denotes the Fresnel propagation kernel. After
propagation through all $S$ slices, the diffraction representation is
computed from the far-field intensity as
\begin{equation}
\mathcal{P}_{\mathrm{EM}}(\mathbf{x}) =
\log_{10}\!\left(
1+
\left\lvert \!\mathcal{F}_{2D}(\psi_S)
\right\rvert^2
\right).
\label{eq:em_diffraction_operator}
\end{equation}
This loss aligns the far-field diffraction patterns of generated and
reference EM volumes.
The overall training objective is
\begin{equation}
\mathcal{L}_{\mathrm{total}} =
\mathcal{L}_{\mathrm{diff}}
+
\lambda_{\mathrm{phys}}\mathcal{L}_{\mathrm{phys}}.
\label{eq:total_training_objective}
\end{equation}
Here, $\lambda_{\mathrm{phys}}$ controls the strength of the domain-specific
regularization; this supervision is applied only during training and does
not alter the inference procedure. The selection of $\lambda_{\mathrm{phys}}$ is analyzed via an ablation over regularization strength in Section~\ref{sec:tuning}.
\section{Results}
\begin{table*}[htbp]
\centering
\caption{
Generalization of compositional conditioning across different diffusion
architectures. Each backbone is evaluated using its original conditioning
strategy and the proposed compositional conditioning. All results are macro-averaged over the 15 imaging modalities.
}
\label{tab:conditioning_comparison}

\setlength{\tabcolsep}{5pt}
\renewcommand{\arraystretch}{1.12}

\begin{tabular}{llccccc}
\toprule
\textbf{Backbone} &
\textbf{Conditioning} &
\multicolumn{2}{c}{\textbf{VMAE}} &
\multicolumn{2}{c}{\textbf{I3D}} &
\textbf{JEDI} $\downarrow$ \\
\cmidrule(lr){3-4}
\cmidrule(lr){5-6}
& &
\textbf{FVD} $\downarrow$ &
\textbf{KID} $(\times 10^{3})$ $\downarrow$ &
\textbf{FVD} $\downarrow$ &
\textbf{KID} $(\times 10^{3})$ $\downarrow$ &
\\
\midrule

\multicolumn{7}{l}{\textbf{Backbone-specific baseline}} \\
\addlinespace[2pt]

MedLoRD \cite{medlord}
& Flattened
& 28.63
& 106.80
& 11.80
& 21.08
& 2.356 \\

Video Diffusion~\cite{VDM2022}
& Original
& 56.87
& 184.97
& 24.00
& 49.28
& 4.63 \\

3DMedDiff~\cite{3dmeddiff}
& Original
& 91.27
& 373.75
& 33.40
& 78.48
& 8.341 \\

Medical Diffusion~\cite{khader2023denoising}
& Original
& 34.55
& 130.12
& 13.19
& 24.08
& \underline{2.732} \\

\midrule

\multicolumn{7}{l}{\textbf{Compositional conditioning}} \\
\addlinespace[2pt]

MedLoRD \cite{medlord}
& Compositional
& \underline{27.30}
& \underline{102.97}
& \underline{11.10}
& \underline{19.69}
& \underline{2.354} \\

Video Diffusion~\cite{VDM2022}
& Compositional
& \underline{51.88}
& \underline{168.03}
& \underline{23.81}
& \underline{48.99}
& \underline{4.40} \\

3DMedDiff~\cite{3dmeddiff}
& Compositional
& \underline{26.85}
& \underline{97.40}
& \underline{11.01}
& \underline{18.69}
& \underline{2.400} \\

Medical Diffusion~\cite{khader2023denoising}
& Compositional
& \underline{33.80}
& \underline{126.91}
& \underline{13.09}
& \underline{23.68}
& 2.733 \\

SciFusion (Ours)
& Compositional
& \textbf{23.51}
& \textbf{84.76}
& \textbf{9.90}
& \textbf{16.59}
& \textbf{1.797} \\

\bottomrule
\end{tabular}

\vspace{2pt}
\begin{minipage}{0.78\textwidth}
\footnotesize
\textit{Note:}
Underline values indicate the better result between baseline and compositional
conditioning for the same backbone.
\end{minipage}
\end{table*}
\subsection{Benchmark Overview}
We construct a volumetric imaging benchmark comprising 15 modalities from three domains: magnetic resonance imaging (MRI), computed tomography (CT), and electron microscopy (EM). The modalities span multiple anatomical structures and imaging contrasts, allowing us to evaluate both unified multimodal generation and compositional generalization across diverse scientific images. EM serves as out-of-distribution data relative to the medical imaging domains, enabling us to assess the framework's robustness and generalizability across fundamentally different imaging modalities. The physical parameters of all datasets are provided in the supplementary
material (Sec.~\ref{supp:phydata}).

\paragraph{Evaluation Metrics.} As each generated sample is a multi-slice
volume, we treat the slice dimension analogously to the temporal
dimension in video and adopt video-based distributional metrics
accordingly, following prior work that applies such metrics to
volumetric medical imaging \cite{hamamci2023generatect}. We evaluate
generation quality against real data using FVD, KID, and JEDI
\cite{unterthiner2019fvd,binkowski2018demystifying,luo2024beyond}.
Specifically, Fréchet Video Distance (FVD) and Kernel Inception Distance
(KID) measure the distributional discrepancy between generated and real
volumes in a pretrained feature space, with lower values indicating closer
alignment to the real data distribution. Since feature extractors trained
on natural video may not fully capture volumetric medical/scientific
structures, we compute FVD and KID using two complementary backbones --
VideoMAE (VMAE) and I3D -- denoted as VMAE-FVD/KID and I3D-FVD/KID,
respectively. JEDI~\cite{luo2024beyond} further provides a joint embedding
distance based on a Joint Embedding Predictive Architecture (JEPA), using
Maximum Mean Discrepancy instead of the Gaussian assumption underlying
FVD/KID; this improves robustness under limited sample sizes, which is
particularly relevant given the number of real samples available for some
modalities.

\subsection{Comparison Across Conditioning Strategies}
\paragraph{Training Data.} For MRI, knee PD, knee PD fat-suppressed, brain T1, brain T2, brain FLAIR, prostate diffusion-weighted, prostate T2-weighted, and breast MRI volumes are obtained from the NYU fastMRI Initiative database~\cite{knoll2020fastmri,zbontar2018fastmri}. For CT, head CT volumes are derived from the RSNA 2019 Brain CT Hemorrhage Challenge dataset~\cite{doi:10.1148/ryai.2020190211}, while chest and abdominal CT volumes are obtained by merging the corresponding
subsets of the Medical Segmentation Decathlon~\cite{simpson2019largeannotatedmedicalimage}. CT data are preprocessed into fixed-depth volumetric samples following the same spatial normalization pipeline used for MRI. The EM domain contains four structural modalities, including cubic and hexagonal crystal structures, apoferritin, and hemoglobin, obtained from the Materials Project~\cite{jain2013commentary} and the Protein Data Bank (PDB)~\cite{Baltruschat2023fRegGANWK}, and processed into volumetric stacks with the same number of slices and spatial resolution as the medical imaging data using \texttt{abTEM}~\cite{madsen2021abtem}. For evaluation, we generate 400 samples per modality and compare the generated distribution against all available real samples of the corresponding modality.

\paragraph{Backbones and Baselines.} Existing approaches typically adopt backbone-specific conditioning mechanisms, where modality information is represented using model-dependent formulations. We investigate whether explicitly factorizing modality information into domain, anatomical object, and imaging contrast components provides a more effective conditioning strategy for multimodal 3D scientific image generation. We evaluate whether compositional conditioning generalizes across diffusion backbones by comparing four representative architectures: MedLoRD~\cite{medlord}, Video Diffusion~\cite{VDM2022}, 3DMedDiff~\cite{3dmeddiff}, and Medical Diffusion~\cite{khader2023denoising}. For MedLoRD, the baseline uses a flattened modality embedding, where each modality is assigned an independent learnable representation. For the remaining backbones, we retain their original conditioning mechanisms as baselines and replace them with the proposed compositional conditioning strategy while keeping all other model components unchanged.
\begin{figure*}[!htb]
    \centering
\includegraphics[width=0.7\textwidth]{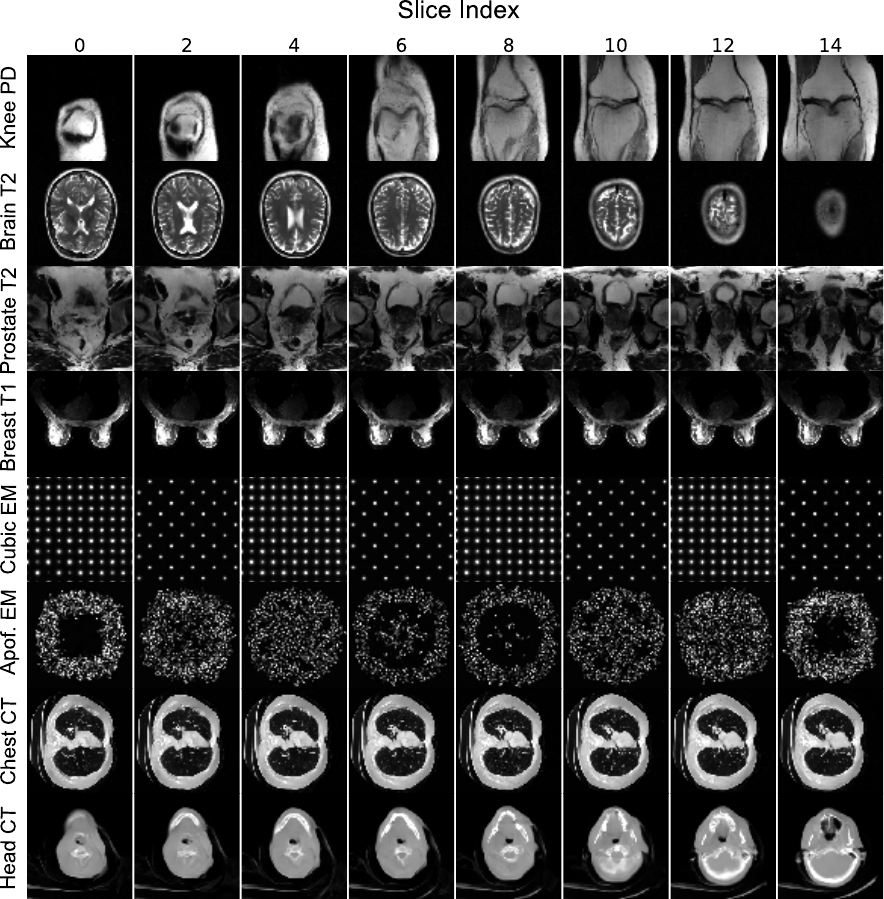}
    \caption{
    Selected generated 3D volumes across all selected modalities.
    }
    \label{fig:generated_selected}
\end{figure*}

As shown in Table~\ref{tab:conditioning_comparison}, compositional conditioning consistently improves generation quality across different diffusion backbones. The proposed representation reduces both VideoMAE-FVD and I3D-FVD for all evaluated architectures, demonstrating improved perceptual fidelity and structural consistency of generated 3D volumes. The improvement is particularly significant for 3DMedDiff, where compositional conditioning reduces VideoMAE-FVD from 91.27 to 26.85 and I3D-FVD from 33.40 to 11.01. 

These results demonstrate that domain--object--contrast factorization provides a generalizable conditioning strategy for multimodal 3D scientific image generation. While Table~\ref{tab:conditioning_comparison} reports results macro-averaged across all 15 modalities, a per-domain (MRI/CT/EM) breakdown is provided in the supplementary material (Table~\ref{tab:conditioning_domain_appendix}).

Notably, SciFusion achieves the best performance across all evaluated metrics, obtaining the lowest VMAE-FVD, VMAE-KID, I3D-FVD, I3D-KID, and JEDI scores among all compared methods. To complement these quantitative results, we further provide qualitative visualizations of the generated volumes. Specifically, we present representative examples from 8 out of the 15 imaging modalities in Fig.~\ref{fig:generated_selected} (see Fig.~\ref{fig:generated_16frames} in supplementary materials for all modalities and slices). The selected modalities span different imaging domains, anatomical structures, and acquisition contrasts, further demonstrating the capability of our unified model to generate diverse 3D scientific images.

\begin{table*}[!htb]
\centering
\caption{Zero-shot compositional generation on unseen target modalities compared
with fully supervised target-modality training and parameter-efficient
adaptation.}
\label{tab:zeroshot_results}
\setlength{\tabcolsep}{5pt}
\renewcommand{\arraystretch}{1.12}
\begin{tabular}{lllccccc}
\toprule
\textbf{Target Modality} &
\textbf{Backbone} &
\textbf{Supervision} &
\multicolumn{2}{c}{\textbf{I3D}} &
\multicolumn{2}{c}{\textbf{VMAE}}
& \textbf{JEDI} $\downarrow$ \\
\cmidrule(lr){4-5}
\cmidrule(lr){6-7}
& & &
\textbf{FVD} $\downarrow$ &
\textbf{KID} $(\times 10^{3})$ $\downarrow$ &
\textbf{FVD} $\downarrow$ &
\textbf{KID} $(\times 10^{3})$ $\downarrow$ \\
\midrule

\multirow{4}{*}{Knee T1}
& SciFusion (Ours)
& Full
& \textbf{13.83}
& \textbf{23.13 $\pm$ 2.00}
& \textbf{14.56}
& \textbf{58.71 $\pm$ 3.78} 
& \textbf{0.35} \\

& SciFusion (Ours)
& Zero-shot
& 26.14
& 50.23 $\pm$ 2.39
& 58.32
& 252.47 $\pm$ 7.07 
& 2.88 \\

& Medical Diffusion \cite{khader2023denoising}
& Zero-shot
& 28.46
& 54.50 $\pm$ 2.65
& 72.46
& 322.46 $\pm$ 7.36 
& 3.59 \\

& VDM + LoRA \cite{VDM2022, hu2022lora}
& PEFT
& 15.99
& 23.06 $\pm$ 1.39
& 25.63
& 102.29 $\pm$ 6.65 
& 1.05 \\

\midrule

\multirow{4}{*}{Brain PD}
& SciFusion (Ours)
& Full
& \textbf{2.90}
& \textbf{4.30 $\pm$ 0.49}
& \textbf{5.22}
& \textbf{19.19 $\pm$ 1.25} 
& \textbf{0.46} \\

& SciFusion (Ours)
& Zero-shot
& 23.52
& 63.84 $\pm$ 3.98
& 15.34
& 64.17 $\pm$ 2.36 
& 1.68 \\

& Medical Diffusion \cite{khader2023denoising}
& Zero-shot
& 11.02
& 23.54 $\pm$ 1.03
& 27.97
& 125.20 $\pm$ 4.35 
& 2.25 \\

& VDM + LoRA \cite{VDM2022, hu2022lora}
& PEFT
& 14.33
& 30.70 $\pm$ 2.00
& 31.13
& 137.24 $\pm$ 4.07
& 1.75 \\
\bottomrule
\end{tabular}
\end{table*}

\subsection{Zero-shot Compositional Generation}
\paragraph{Zero-shot Setup.} To evaluate compositional zero-shot generation, we consider two unseen modalities whose factor combinations are excluded from training. Knee T1-weighted MRI is obtained from MRNet~\cite{stanford_mrnet_dataset, bien2018mrnet}, while brain PD-weighted MRI is obtained from the IXI dataset~\cite{ixi_dataset}. Both modalities combine domains, anatomical objects, and contrasts that are individually observed during training but never presented together as complete modality combinations, providing a direct test of whether the proposed compositional conditioning mechanism can recombine learned semantic factors to generate unseen modalities. We generate 600 samples per target modality and compare SciFusion's zero-shot generation against two baselines: fully supervised single-modality diffusion models trained directly on target-modality samples, which represent an upper bound with direct access to the target distribution; and parameter-efficient adaptation (PEFT) using LoRA~\cite{hu2022lora}, which adapts a pretrained model from a related observed modality to the target modality (Knee PD~$\rightarrow$~Knee T1, Brain T1~$\rightarrow$~Brain PD) using additional target-modality samples.




During training, the model observes no samples from the target modalities; each target is instead synthesized at inference time by recomposing previously learned domain, object, and contrast representations (e.g., combining a learned knee/brain object with a learned T1/PD contrast), without additional optimization.

As shown in Table~\ref{tab:zeroshot_results}, fully supervised models achieve the best performance due to direct access to target-modality samples, while the gap on zero-shot targets reflects the inherent difficulty of synthesizing unseen imaging contrasts without target-specific data. Nonetheless, the proposed compositional model generates unseen modalities without observing any target-modality samples or performing additional optimization, indicating that the learned compositional representation captures transferable relationships between anatomical structures and imaging contrasts. PEFT narrows this gap on knee T1 by adapting the pretrained model with target-modality samples and additional optimization; notably, on brain PD our training-free zero-shot generation already matches or exceeds PEFT on JEDI (1.68 vs.\ 1.75), suggesting that the learned composition for this target is already well aligned without adaptation.

\begin{figure}[!htb]
\centering
\includegraphics[width=0.45\textwidth]{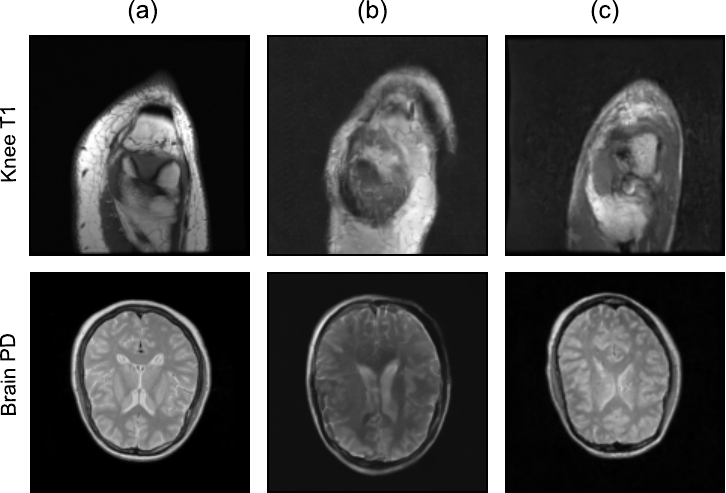}
\caption{
Qualitative comparison of generation on unseen target modalities (knee T1 and
brain PD). 
(a) Real target-modality images, 
(b) our zero-shot generation without access to target-modality data or additional
training, and 
(c) parameter-efficient fine-tuning (PEFT) using VDM with LoRA and
target-modality samples.
}
    \label{fig:zero_shot}
\end{figure}

Qualitative comparisons are provided in Fig.~\ref{fig:zero_shot}. The zero-shot generation produces visually plausible anatomical structures and modality-specific appearance without accessing any target-modality samples. These qualitative results suggest the potential of compositional conditioning for synthesizing unseen imaging modalities in a training-free manner.

\begin{table*}[!htb]
\centering
\caption{Ablation study of SciFusion across four configurations: single-modality models, flattened multimodal conditioning, compositional conditioning, and the full model with physics guidance.}
\label{tab:single_unified_ablation}

\setlength{\tabcolsep}{7pt}
\renewcommand{\arraystretch}{1.15}

\begin{tabular}{lcccccc}
\hline
\textbf{Method}
& \textbf{\# Models}
& \textbf{Comp.}
& \textbf{Physics}
& \textbf{I3D-FVD $\downarrow$}
& \textbf{VMAE-FVD $\downarrow$}
& \textbf{JEDI $\downarrow$} \\
\hline

Single Modality
& 15
& --
& --
& 13.07
& 35.93
& 2.62 \\

\hline

Flat Multimodal
& 1
& --
& --
& 11.80
& 28.63
& 2.36 \\

Comp. Multimodal
& 1
& \checkmark
& --
& 11.10
& 27.30
& 2.35 \\

Ours
& 1
& \checkmark
& \checkmark
& \textbf{9.90}
& \textbf{23.51}
& \textbf{1.80} \\

\hline
\end{tabular}
\end{table*}

\subsection{Ablation Study}
\paragraph{Unified Multimodal Generation.}

We further perform an ablation study to investigate the contribution of
different conditioning strategies and physical guidance proposed in SciFusion. 
These comparisons disentangle the effects of parameter
sharing, semantic decomposition of modality information, and physical priors
on multimodal generation. As shown in Table~\ref{tab:single_unified_ablation}, the unified multimodal
model achieves comparable or better generation quality while reducing the
number of models from 15 to 1. The flattened multimodal model improves all evaluation
metrics, demonstrating that joint training across heterogeneous modalities can effectively exploit shared information between domains, compared
with training a separate model for each single modality. Introducing compositional  conditioning further
improves generation quality, reducing I3D-FVD from 11.80 to 11.10 and
VideoMAE-FVD from 28.63 to 27.30, indicating that modeling the underlying semantic factors of scientific
imaging modalities provides a more effective conditioning representation
than treating each modality as an independent category. Finally, adding physics-guided regularization achieves the best overall
performance, with improvements across all evaluation metrics. These results
demonstrate that unified multimodal learning, compositional conditioning, and
physical constraints provide complementary benefits for scalable 3D scientific imaging.

\paragraph{Physical Loss Weight Sensitivity.}
\label{sec:tuning}
We further investigate the sensitivity of the physical-loss weight
$\lambda_{\mathrm{phy}}$ across three values ($2\cdot10^{-4}$,
$4\cdot10^{-4}$, $5\cdot10^{-4}$), keeping all other training
configurations unchanged and evaluating at approximately the same
training stage. Table~\ref{tab:physical_scale_ablation} shows that
$\lambda_{\mathrm{phy}}=2\cdot10^{-4}$ consistently underperforms across
all domains, indicating a minimum regularization strength is needed for
the constraints to take effect. For MRI and CT, $5\cdot10^{-4}$ is
uniformly best across all metrics; for EM, $4\cdot10^{-4}$ gives better
FVD/KID but $5\cdot10^{-4}$ gives substantially better JEDI (1.50 vs.\
2.13). Averaged across all three domains, $5\cdot10^{-4}$ achieves the
best overall performance on every metric, so we adopt it as a single
default weight across domains, consistent with our shared training
configuration.

\begin{table}[!htb] \centering \caption{Effect of physical-loss regularization strength across different imaging
domains. KID is multiplied by $10^3$ for readability.} \label{tab:physical_scale_ablation} \setlength{\tabcolsep}{2.5pt} \renewcommand{\arraystretch}{1.12} \begin{tabular}{llccccc} \toprule \textbf{Domain} & $\boldsymbol{\lambda_{\mathrm{phy}}}$ & \multicolumn{2}{c}{\textbf{I3D}} & \multicolumn{2}{c}{\textbf{VMAE}} & \textbf{JEDI} $\downarrow$ \\ \cmidrule(lr){3-4} \cmidrule(lr){5-6} & & \textbf{FVD} $\downarrow$ & \textbf{KID} $\downarrow$ & \textbf{FVD} $\downarrow$ & \textbf{KID} $\downarrow$ & \\ \midrule \multirow{3}{*}{MRI} & $5\cdot10^{-4}$ & \textbf{10.18} & \textbf{18.04} & \textbf{24.70} & \textbf{95.35} & \textbf{1.46} \\ & $4\cdot10^{-4}$ & 11.59 & 22.26 & 28.69 & 114.18 & 1.64 \\ & $2\cdot10^{-4}$ & 11.46 & 21.40 & 28.95 & 115.29 & 1.64 \\ \midrule \multirow{3}{*}{EM} & $5\cdot10^{-4}$ & 5.42 & 6.28 & 16.85 & 46.68 & \textbf{1.50} \\ & $4\cdot10^{-4}$ & \textbf{5.04} & \textbf{5.59} & \textbf{15.15} & \textbf{38.46} & 2.13 \\ & $2\cdot10^{-4}$ & 5.87 & 7.51 & 18.06 & 50.54 & 2.44 \\ \midrule \multirow{3}{*}{CT} & $5\cdot10^{-4}$ & \textbf{16.72} & \textbf{30.38} & \textbf{31.81} & \textbf{123.55} & \textbf{3.08} \\ & $4\cdot10^{-4}$ & 21.89 & 45.18 & 36.66 & 142.38 & 3.37 \\ & $2\cdot10^{-4}$ & 17.83 & 33.58 & 34.92 & 136.25 & 3.45 \\ \midrule \multirow{3}{*}{\textbf{Overall}} & $5\cdot10^{-4}$ & \textbf{9.90} & \textbf{16.59} & \textbf{23.51} & \textbf{84.77} & \textbf{1.80} \\ & $4\cdot10^{-4}$ & 11.47 & 21.29 & 25.77 & 94.58 & 2.15 \\ & $2\cdot10^{-4}$ & 10.87 & 19.20 & 26.51 & 97.90 & 2.27 \\ \bottomrule \end{tabular} \end{table}
\section{Conclusion}

Scientific imaging often faces a practical imbalance in data availability: while some imaging domains and acquisition settings are well represented, newly introduced contrasts, protocols, or imaging configurations may have only limited data. We presented SciFusion, which mitigates this challenge by exploiting shared structure across modalities through compositional conditioning, while domain-specific physics objectives provide additional guidance tailored to different image formation processes.

Across 15 medical and scientific imaging modalities, compositional conditioning and physics guidance together improve generation quality over conventional conditioning strategies, reducing I3D-FVD from 11.80 to 9.90 relative to a flattened multimodal baseline, with both components contributing to the improvement. Beyond improved generation fidelity, factorizing modality information into reusable semantic components allows the model to synthesize previously unseen combinations of known factors without training on the target composition. This setting is particularly relevant when a new imaging configuration shares characteristics with existing modalities—for example, an established anatomical target acquired under a new contrast or protocol. In such cases, knowledge learned from related data can be reused before a large target-specific dataset becomes available,  allowing generated samples to serve as an initial data source for method development while a complete target-specific dataset is collected.
\paragraph{Limitations and Future Work}
The effectiveness of zero-shot compositional transfer depends on the extent to which a target modality can be represented by factors already observed during training. When the target exhibits substantial acquisition-specific or distributional differences, recombining known factors alone may be insufficient to capture its characteristics. A natural extension is therefore to combine compositional transfer with lightweight target-specific guidance or few-shot adaptation when limited real samples are available. Future work will also explore automatically learned compositional representations beyond manually defined categories and evaluate the approach across broader imaging domains, scanners, and acquisition protocols.

\newpage
{
    \small
    \bibliographystyle{ieeenat_fullname}
    \bibliography{main}

@String(CVPR= {IEEE Conf. Comput. Vis. Pattern Recog.})

@String(ECCV= {Eur. Conf. Comput. Vis.})

@String(NIPS= {Adv. Neural Inform. Process. Syst.})

@String(ICASSP=	{ICASSP})

@String(ICLR = {Int. Conf. Learn. Represent.})

@String(AAAI = {AAAI})

@String(CVPR  = {CVPR})

@String(ECCV  = {ECCV})

@String(NIPS  = {NeurIPS})

@String(ICLR  = {ICLR})

@inproceedings{Bangun_2025,
   title={MRI Reconstruction with Regularized 3D Diffusion Model (R3DM)},
   url={http://dx.doi.org/10.1109/WACV61041.2025.00078},
   DOI={10.1109/wacv61041.2025.00078},
   booktitle={2025 IEEE/CVF Winter Conference on Applications of Computer Vision (WACV)},
   publisher={IEEE},
   author={Bangun, Arya and Cao, Zhuo and Quercia, Alessio and Scharr, Hanno and Pfaehler, Elisabeth},
   year={2025},
   month=Feb, pages={700–710} }

@article{tudosiu2024realistic,
  title   = {Realistic morphology-preserving generative modelling of the brain},
  author  = {Tudosiu, Petru-Daniel and Pinaya, Walter Hugo Lopez and Ferreira Da Costa, Pedro and others},
  journal = {Nature Machine Intelligence},
  volume  = {6},
  pages   = {811--819},
  year    = {2024},
  doi     = {10.1038/s42256-024-00864-0}
}

@article{khader2023denoising,
  title   = {Denoising diffusion probabilistic models for {3D} medical image generation},
  author  = {Khader, Firas and M{\"u}ller-Franzes, Gustav and
             Tayebi Arasteh, Soroosh and others},
  journal = {Scientific Reports},
  volume  = {13},
  number  = {1},
  pages   = {7303},
  year    = {2023},
  doi     = {10.1038/s41598-023-34341-2}
}

@InProceedings{medlord,
author="Seyfarth, Marvin
and Dar, Salman Ul Hassan
and Ayx, Isabelle
and Fink, Matthias Alexander
and Schoenberg, Stefan O.
and Kauczor, Hans-Ulrich
and Engelhardt, Sandy",
editor="Fernandez, Virginia
and Wiesner, David
and Zuo, Lianrui
and Casamitjana, Adri{\`a}
and Remedios, Samuel W.",
title="MedLoRD: A Medical Low-Resource Diffusion Model for High-Resolution 3D CT Image Synthesis",
booktitle="Simulation and Synthesis in Medical Imaging",
year="2026",
publisher="Springer Nature Switzerland",
address="Cham",
pages="1--12",
isbn="978-3-032-05573-6"
}

@article{3dmeddiff,
  title={3D MedDiffusion: A 3D Medical Latent Diffusion Model for Controllable and High-Quality Medical Image Generation},
  author={Haoshen Wang and Zhentao Liu and Kaicong Sun and Xiaodong Wang and Dinggang Shen and Zhiming Cui},
  journal={IEEE Transactions on Medical Imaging},
  year={2024},
  volume={44},
  pages={4960-4972},
  url={https://api.semanticscholar.org/CorpusID:274789446}
}

@article{Chen2024TowardsGT,
  title={Towards Generalizable Tumor Synthesis},
  author={Qi Chen and Xiaoxi Chen and Haorui Song and Zhiwei Xiong and Alan L. Yuille and Chen Wei and Zongwei Zhou},
  journal={2024 IEEE/CVF Conference on Computer Vision and Pattern Recognition (CVPR)},
  year={2024},
  pages={11147-11158},
  url={https://api.semanticscholar.org/CorpusID:268091302}
}

@article{maisi-v2, title={MAISI-v2: Accelerated 3D High-Resolution Medical Image Synthesis with Rectified Flow and Region-specific Contrastive Loss}, volume={40}, url={https://ojs.aaai.org/index.php/AAAI/article/view/38309}, DOI={10.1609/aaai.v40i15.38309}, abstractNote={Medical image synthesis is an important topic for both clinical and research applications. Recently, diffusion models have become a leading approach in this area. Despite their strengths, many existing methods struggle with (1) limited generalizability, only working for specific body regions or voxel spacings, (2) slow inference, which is a common issue for diffusion models, and (3) weak alignment with input conditions, which is a critical issue for medical imaging. MAISI, a previously proposed framework, addresses generalizability issues but still suffers from slow inference and limited condition consistency. In this work, we present MAISI-v2, the first accelerated 3D medical image synthesis framework that integrates rectified flow to enable fast and high-quality generation. To further enhance condition fidelity, we introduce a novel region-specific contrastive loss to improve sensitivity to the region of interest. Our experiments show that MAISI-v2 can achieve state-of-the-art image quality with 33× acceleration for latent diffusion models. We also conducted a downstream segmentation experiment to show that the synthetic images can be used for data augmentation. We release our code, training details, model weights, and a GUI demo to facilitate reproducibility and promote further development within the community.}, number={15}, journal={Proceedings of the AAAI Conference on Artificial Intelligence}, author={Zhao, Can and Guo, Pengfei and Yang, Dong and He, Yufan and Tang, Yucheng and Simon, Benjamin and Belue, Mason and Harmon, Stephanie and Turkbey, Baris and Xu, Daguang}, year={2026}, month={Mar.}, pages={13088–13098} }

@inproceedings{physicsinformedldm,
author = {L\"{u}pke, Sven and Yeganeh, Yousef and Adeli, Ehsan and Navab, Nassir and Farshad, Azade},
title = {Physics-Informed Latent Diffusion for\&nbsp;Multimodal Brain MRI Synthesis},
year = {2024},
isbn = {978-3-031-84524-6},
publisher = {Springer-Verlag},
address = {Berlin, Heidelberg},
url = {https://doi.org/10.1007/978-3-031-84525-3_17},
doi = {10.1007/978-3-031-84525-3_17},
booktitle = {Medical Image Computing and Computer Assisted Intervention – MICCAI 2024 Workshops: LDTM 2024, MMMI/ML4MHD 2024, ML-CDS 2024, Held in Conjunction with MICCAI 2024, Marrakesh, Morocco, October 6–10, 2024, Proceedings},
pages = {198–207},
numpages = {10},
location = {Marrakesh, Morocco}
}

@INPROCEEDINGS{maisi-v1,
  author={Guo, Pengfei and Zhao, Can and Yang, Dong and Xu, Ziyue and Nath, Vishwesh and Tang, Yucheng and Simon, Benjamin and Belue, Mason and Harmon, Stephanie and Turkbey, Baris and Xu, Daguang},
  booktitle={2025 IEEE/CVF Winter Conference on Applications of Computer Vision (WACV)}, 
  title={MAISI: Medical AI for Synthetic Imaging}, 
  year={2025},
  volume={},
  number={},
  pages={4430-4441},
  doi={10.1109/WACV61041.2025.00435}}

@misc{blattmann2023alignlatentshighresolutionvideo,
      title={Align your Latents: High-Resolution Video Synthesis with Latent Diffusion Models}, 
      author={Andreas Blattmann and Robin Rombach and Huan Ling and Tim Dockhorn and Seung Wook Kim and Sanja Fidler and Karsten Kreis},
      year={2023},
      eprint={2304.08818},
      archivePrefix={arXiv},
      primaryClass={cs.CV},
      url={https://arxiv.org/abs/2304.08818}, 
}

@ARTICLE{HeRadonInversion,
  author={He, Ji and Wang, Yongbo and Ma, Jianhua},
  journal={IEEE Transactions on Medical Imaging}, 
  title={Radon Inversion via Deep Learning}, 
  year={2020},
  volume={39},
  number={6},
  pages={2076-2087},
  doi={10.1109/TMI.2020.2964266}}

@article{xu2025DeepRadon,
author = {Xu, Shuo and Fu, Jintao and Sun, Yuewen and Cong, Peng and Xiang, Xincheng},
title = {Deep Radon Prior: A fully unsupervised framework for sparse-view CT reconstruction},
year = {2025},
issue_date = {May 2025},
publisher = {Pergamon Press, Inc.},
address = {USA},
volume = {189},
number = {C},
issn = {0010-4825},
url = {https://doi.org/10.1016/j.compbiomed.2025.109853},
doi = {10.1016/j.compbiomed.2025.109853},
journal = {Comput. Biol. Med.},
month = may,
numpages = {18}
}

@article{Baltruschat2023fRegGANWK,
  title={fRegGAN with K-space Loss Regularization for Medical Image Translation},
  author={Ivo M. Baltruschat and Felix Kreis and Alexander Hoelscher and Melanie Dohmen and Matthias Lenga},
  journal={ArXiv},
  year={2023},
  volume={abs/2303.15938},
  url={https://api.semanticscholar.org/CorpusID:257771652}
}

@INPROCEEDINGS {ldm2022,
author = { Rombach, Robin and Blattmann, Andreas and Lorenz, Dominik and Esser, Patrick and Ommer, Bjorn },
booktitle = { 2022 IEEE/CVF Conference on Computer Vision and Pattern Recognition (CVPR) },
title = {{ High-Resolution Image Synthesis with Latent Diffusion Models }},
year = {2022},
volume = {},
ISSN = {},
pages = {10674-10685},
doi = {10.1109/CVPR52688.2022.01042},
url = {https://doi.ieeecomputersociety.org/10.1109/CVPR52688.2022.01042},
publisher = {IEEE Computer Society},
address = {Los Alamitos, CA, USA},
month =Jun}

@misc{stanford_mrnet_dataset,
  author       = {{Stanford Center for Artificial Intelligence in Medicine and Imaging}},
  title        = {{MRNet: Knee MRIs}},
  howpublished = {\url{https://doi.org/10.71718/rcbp-8c35}},
  doi          = {10.71718/rcbp-8c35},
  note         = {Accessed: 2026-08-04}
}

@article{bien2018mrnet,
  author  = {Bien, Nicholas and Rajpurkar, Pranav and Ball, Robyn L. and
             Irvin, Jeremy and Park, Allison and Jones, Erik and
             Bereket, Michael and Patel, Bhavik N. and Yeom, Kristen W. and
             Shpanskaya, Katie and Halabi, Safwan and Zucker, Evan and
             Fanton, Gary and Amanatullah, Derek F. and Beaulieu, Christopher F.
             and Riley, Geoffrey M. and Stewart, Robert J. and Blankenberg,
             Francis G. and Larson, David B. and Jones, Richard H. and
             Langlotz, Curtis P. and Ng, Andrew Y. and Lungren, Matthew P.},
  title   = {Deep-learning-assisted diagnosis for knee magnetic resonance
             imaging: Development and retrospective validation of MRNet},
  journal = {PLOS Medicine},
  year    = {2018},
  volume  = {15},
  number  = {11},
  pages   = {e1002699},
  doi     = {10.1371/journal.pmed.1002699}
}

@article{knoll2020fastmri,
  title     = {{fastMRI}: A Publicly Available Raw {k}-Space and {DICOM}
               Dataset of Knee Images for Accelerated {MR} Image
               Reconstruction Using Machine Learning},
  author    = {Knoll, Florian and Zbontar, Jure and Sriram, Anuroop
               and Muckley, Matthew J. and Bruno, Mary and Defazio, Aaron
               and Parente, Marc and Geras, Krzysztof J. and Katsnelson, Joe
               and Chandarana, Hersh and Zhang, Zizhao and Drozdzal, Michal
               and Romero, Adriana and Rabbat, Michael and Vincent, Pascal
               and Pinkerton, James and Wang, Duo and Yakubova, Nafissa
               and Owens, Erich and Zitnick, C. Lawrence and Recht, Michael P.
               and Sodickson, Daniel K. and Lui, Yvonne W.},
  journal   = {Radiology: Artificial Intelligence},
  volume    = {2},
  number    = {1},
  pages     = {e190007},
  year      = {2020},
  doi       = {10.1148/ryai.2020190007}
}

@article{zbontar2018fastmri,
  title         = {{fastMRI}: An Open Dataset and Benchmarks for
                   Accelerated {MRI}},
  author        = {Zbontar, Jure and Knoll, Florian and Sriram, Anuroop
                   and Murrell, Tullie and Huang, Zhengnan
                   and Muckley, Matthew J. and Defazio, Aaron
                   and Stern, Ruben and Johnson, Patricia and Bruno, Mary
                   and Parente, Marc and Geras, Krzysztof J.
                   and Katsnelson, Joe and Chandarana, Hersh
                   and Zhang, Zizhao and Drozdzal, Michal
                   and Romero, Adriana and Rabbat, Michael
                   and Vincent, Pascal and Yakubova, Nafissa
                   and Pinkerton, James and Wang, Duo and Owens, Erich
                   and Zitnick, C. Lawrence and Recht, Michael P.
                   and Sodickson, Daniel K. and Lui, Yvonne W.},
  journal       = {arXiv preprint arXiv:1811.08839},
  year          = {2018},
  eprint        = {1811.08839},
  archivePrefix = {arXiv},
  primaryClass  = {cs.CV}
}

@inproceedings{VDM2022,
author = {Ho, Jonathan and Salimans, Tim and Gritsenko, Alexey and Chan, William and Norouzi, Mohammad and Fleet, David J.},
title = {Video diffusion models},
year = {2022},
isbn = {9781713871088},
publisher = {Curran Associates Inc.},
address = {Red Hook, NY, USA},
booktitle = {Proceedings of the 36th International Conference on Neural Information Processing Systems},
articleno = {628},
numpages = {14},
location = {New Orleans, LA, USA},
series = {NIPS '22}
}

@misc{simpson2019largeannotatedmedicalimage,
      title={A large annotated medical image dataset for the development and evaluation of segmentation algorithms}, 
      author={Amber L. Simpson and Michela Antonelli and Spyridon Bakas and Michel Bilello and Keyvan Farahani and Bram van Ginneken and Annette Kopp-Schneider and Bennett A. Landman and Geert Litjens and Bjoern Menze and Olaf Ronneberger and Ronald M. Summers and Patrick Bilic and Patrick F. Christ and Richard K. G. Do and Marc Gollub and Jennifer Golia-Pernicka and Stephan H. Heckers and William R. Jarnagin and Maureen K. McHugo and Sandy Napel and Eugene Vorontsov and Lena Maier-Hein and M. Jorge Cardoso},
      year={2019},
      eprint={1902.09063},
      archivePrefix={arXiv},
      primaryClass={cs.CV},
      url={https://arxiv.org/abs/1902.09063}, 
}

@article{doi:10.1148/ryai.2020190211,
author = {Flanders, Adam                             E. and Prevedello, Luciano                             M. and Shih, George and Halabi, Safwan                             S. and Kalpathy-Cramer, Jayashree and Ball, Robyn and Mongan, John                         T. and Stein, Anouk and Kitamura, Felipe                             C. and Lungren, Matthew                             P. and Choudhary, Gagandeep and Cala, Lesley and Coelho, Luiz and Mogensen, Monique and Mor\'{o}n, Fanny and Miller, Elka and Ikuta, Ichiro and Zohrabian, Vahe and McDonnell, Olivia and Lincoln, Christie and Shah, Lubdha and Joyner, David and Agarwal, Amit and Lee, Ryan                         K. and Nath, Jaya and },
title = {Construction of a Machine Learning Dataset through Collaboration: The                     RSNA 2019 Brain CT Hemorrhage Challenge},
journal = {Radiology: Artificial Intelligence},
volume = {2},
number = {3},
pages = {e190211},
year = {2020},
doi = {10.1148/ryai.2020190211},
    note ={PMID: 33937827},

URL = { 
    
        https://doi.org/10.1148/ryai.2020190211
    
    

},
eprint = { 
    
        https://doi.org/10.1148/ryai.2020190211
    
    

}

}

@inproceedings{ddpm2020,
author = {Ho, Jonathan and Jain, Ajay and Abbeel, Pieter},
title = {Denoising diffusion probabilistic models},
year = {2020},
isbn = {9781713829546},
publisher = {Curran Associates Inc.},
address = {Red Hook, NY, USA},
booktitle = {Proceedings of the 34th International Conference on Neural Information Processing Systems},
articleno = {574},
numpages = {12},
location = {Vancouver, BC, Canada},
series = {NIPS '20}
}

@inproceedings{valdy2026physics,
  title={Physics-Guided Diffusion Priors for Multi-Slice Reconstruction in Scientific Imaging},
  author={Valdy, Laurentius and Paul, Richard D and Quercia, Alessio and Cao, Zhuo and Zhao, Xuan and Scharr, Hanno and Bangun, Arya},
  booktitle={ICASSP 2026-2026 IEEE International Conference on Acoustics, Speech and Signal Processing (ICASSP)},
  pages={6271--6275},
  year={2026},
  organization={IEEE}
}

@article{jain2013commentary,
  title={Commentary: The Materials Project: A materials genome approach to accelerating materials innovation},
  author={Jain, Anubhav and Ong, Shyue Ping and Hautier, Geoffroy and Chen, Wei and Richards, William Davidson and Dacek, Stephen and Cholia, Shreyas and Gunter, Dan and Skinner, David and Ceder, Gerbrand and others},
  journal={APL materials},
  volume={1},
  number={1},
  year={2013},
  publisher={AIP Publishing}
}

@article{madsen2021abtem,
  title={The abTEM code: transmission electron microscopy from first principles},
  author={Madsen, Jacob and Susi, Toma},
  journal={Open Research Europe},
  volume={1},
  pages={24},
  year={2021}
}

@book{kirkland1998advanced,
  title={Advanced computing in electron microscopy},
  author={Kirkland, Earl J},
  volume={12},
  year={1998},
  publisher={Springer}
}

@article{cowley1957scattering,
  title={The scattering of electrons by atoms and crystals. I. A new theoretical approach},
  author={Cowley, John M and Moodie, Alexander Forbes},
  journal={Acta Crystallographica},
  volume={10},
  number={10},
  pages={609--619},
  year={1957},
  publisher={International Union of Crystallography}
}

@article{cowley1959scattering,
  title={The scattering of electrons by atoms and crystals. II. The effects of finite source size},
  author={Cowley, JM and Moodie, AF},
  journal={Acta Crystallographica},
  volume={12},
  number={5},
  pages={353--359},
  year={1959},
  publisher={International Union of Crystallography}
}

@article{cowley1959scattering2,
  title={The scattering of electrons by atoms and crystals. III. Single-crystal diffraction patterns},
  author={Cowley, JM t and Moodie, AF},
  journal={Acta Crystallographica},
  volume={12},
  number={5},
  pages={360--367},
  year={1959},
  publisher={International Union of Crystallography}
}

@article{ishizuka1977new,
  title={A new theoretical and practical approach to the multislice method},
  author={Ishizuka, KAZUO and Uyeda, NATSU},
  journal={Foundations of Crystallography},
  volume={33},
  number={5},
  pages={740--749},
  year={1977},
  publisher={International Union of Crystallography}
}

@article{lobato2014accurate,
  title={An accurate parameterization for scattering factors, electron densities and electrostatic potentials for neutral atoms that obey all physical constraints},
  author={Lobato, I and Van Dyck, D},
  journal={Foundations of Crystallography},
  volume={70},
  number={6},
  pages={636--649},
  year={2014},
  publisher={International Union of Crystallography}
}

@misc{ixi_dataset,
  author       = {{Information eXtraction from Images Project}},
  title        = {{IXI Dataset}},
  howpublished = {\url{https://brain-development.org/ixi-dataset/}},
  institution  = {Imperial College London},
  note         = {Licensed under CC BY-SA 3.0; accessed August 4, 2026}
}

@article{song2020score,
  title={Score-based generative modeling through stochastic differential equations},
  author={Song, Yang and Sohl-Dickstein, Jascha and Kingma, Diederik P and Kumar, Abhishek and Ermon, Stefano and Poole, Ben},
  journal={arXiv preprint arXiv:2011.13456},
  year={2020}
}

@article{chung2022score,
  title={Score-based diffusion models for accelerated MRI},
  author={Chung, Hyungjin and Ye, Jong Chul},
  journal={Medical image analysis},
  volume={80},
  pages={102479},
  year={2022},
  publisher={Elsevier}
}

@inproceedings{
  song2021scorebased,
  title={Score-Based Generative Modeling through Stochastic Differential Equations},
  author={Yang Song and Jascha Sohl-Dickstein and Diederik P Kingma and Abhishek Kumar and Stefano Ermon and Ben Poole},
  booktitle={International Conference on Learning Representations},
  year={2021},
  url={https://openreview.net/forum?id=PxTIG12RRHS}
}

@article{luo2024beyond,
  title={Beyond {FVD}: Enhanced Evaluation Metrics for Video Generation Quality},
  author={Luo, Ge Ya and Favero, Gian Mario and Luo, Zhi Hao and Jolicoeur-Martineau, Alexia and Pal, Christopher},
  journal = {ICLR},
  year={2025},
}

@article{unterthiner2019fvd,
  title = {{FVD}: A new metric for video generation},
  author= {Unterthiner, Thomas and van Steenkiste, Sjoerd and Kurach, Karol and Marinier, Rapha{\"e}l and Michalski, Marcin and Gelly, Sylvain},
  journal = {ICLR Workshop DeepGenStruct},
  year = {2019},
}

@inproceedings{
hu2022lora,
title={Lo{RA}: Low-Rank Adaptation of Large Language Models},
author={Edward J Hu and Yelong Shen and Phillip Wallis and Zeyuan Allen-Zhu and Yuanzhi Li and Shean Wang and Lu Wang and Weizhu Chen},
booktitle={International Conference on Learning Representations},
year={2022},
url={https://openreview.net/forum?id=nZeVKeeFYf9}
}

@article{binkowski2018demystifying,
  title={Demystifying mmd gans},
  author={Bi{\'n}kowski, Miko{\l}aj and Sutherland, Danica J and Arbel, Michael and Gretton, Arthur},
  journal={International Conference on Learning Representations},
  year={2018}
}

@InProceedings{hamamci2023generatect,
author="Hamamci, Ibrahim Ethem
and Er, Sezgin
and Sekuboyina, Anjany
and Simsar, Enis
and Tezcan, Alperen
and Simsek, Ayse Gulnihan
and Esirgun, Sevval Nil
and Almas, Furkan
and Do{\u{g}}an, Irem
and Dasdelen, Muhammed Furkan
and Prabhakar, Chinmay
and Reynaud, Hadrien
and Pati, Sarthak
and Bluethgen, Christian
and Ozdemir, Mehmet Kemal
and Menze, Bjoern",
editor="Leonardis, Ale{\v{s}}
and Ricci, Elisa
and Roth, Stefan
and Russakovsky, Olga
and Sattler, Torsten
and Varol, G{\"u}l",
title="GenerateCT: Text-Conditional Generation of 3D Chest CT Volumes",
booktitle="Computer Vision -- ECCV 2024",
year="2025",
publisher="Springer Nature Switzerland",
address="Cham",
pages="126--143",
isbn="978-3-031-72986-7"
}
}

\newpage

\appendix
\section*{Supplementary Materials}
\section{Dataset and Preprocessing Details}
\label{supp:phydata}
\paragraph{Magnetic Resonance Imaging (MRI).}
We collect volumetric data from fastMRI dataset \cite{zbontar2018fastmri}, Brain IXI \cite{ixi_dataset}, Stanford Knee \cite{stanford_mrnet_dataset, bien2018mrnet}. The datasets are acquired with the field strength of $1.5$ T and $3$ T. All data are pre-processed to have depth, height, and width dimension $16 \times 256 \times 256$. The slice dimension is chosen such that we have important feature.  The Brain IXI and Stanford Knee data are used only to assess the zero shot generation using compositional embedding, hence, we do not use the data for training the model.  The summary is given in Table \ref{tab:mri_datasets}.
\begin{table}[htbp]
\centering
\small 
\caption{Summary of MRI datasets}
\label{tab:mri_datasets}
\renewcommand{\arraystretch}{1.2}
\begin{tabular}{llll}
\textbf{Dataset} &
\textbf{Anatomy} & 
\textbf{Contrast} & \textbf{Total} \\
\hline
fastMRI &
Knee &
PD and PDFS & (480, 486)\\
fastMRI &
Brain & 
T1,T2, and FLAIR & (328, 1367, 2547)\\

fastMRI &
Prostate & 
T2 and DWI & (272, 261) \\

fastMRI &
Breast & 
T1 & 609\\

IXI &
Brain &  PD  & 566\\

MRNet &
Knee &  T1 &1130 \\

\hline
\end{tabular}
\end{table}
\paragraph{Computed Tomography (CT).} We collect CT data from RSNA Brain data \cite{doi:10.1148/ryai.2020190211} and Medical Segmentation Decathlon \cite{simpson2019largeannotatedmedicalimage}. The images are pre-processed such that we have $256 \times 256$ with the number of slices $16$. The summary is given in Table \ref{tab:ct_datasets}.

\begin{table}[htbp]
\centering
\small
\caption{Summary of the CT datasets}
\label{tab:ct_datasets}
\renewcommand{\arraystretch}{1.2}

\begin{tabular}{p{2cm} p{2cm} p{2.5cm}}
\toprule
\textbf{Dataset} &
\textbf{Anatomy} &
\textbf{Total}  \\
\midrule

RSNA&
Brain &
339 exams (subset)   \\

MSD &
Liver &
201 scans  \\

MSD &
Lung &
96 scans  \\

MSD&
Pancreas &
420 scans  \\

MSD&
Colon &
190 scans  \\

MSD&
Liver &
443 scans  \\

MSD&
Spleen &
61 scans   \\

\bottomrule
\end{tabular}
\end{table}
\paragraph{Electron Microscopy (EM).}
Table~\ref{tab:abtem_params} summarizes the abTEM simulation parameters for both crystalline and protein datasets, including cell size, spatial resolution, slice discretization, and propagation settings.  In both cases, simulations are performed using plane-wave illumination. For crystal materials we construct cubic and hexagonal structure with total volumetric data $278$ and $160$, respectively. Additonally for protein we collect $69$ apoferritin and $204$ hemoglobin data. 
\begin{table}[!htb]
\centering
\small
\caption{abTEM simulation parameters used for dataset generation.}
\begin{tabular}{lcc}
\hline
Parameter & Crystal data & Protein data \\
\hline
Pixel Dimension & $ 256 \times 256 $ & $ 256 \times 256$ \\
Sampling / pixel size & $0.1~\text{\AA}$ & $1.0~\text{\AA}$ \\ 
Number of slices $(S)$ & $16$ & $16$ \\
Slice thickness & $0.5~\text{\AA}$ & $4.0~\text{\AA}$ \\
Potential & Lobato \cite{lobato2014accurate} & Lobato\cite{lobato2014accurate} \\ 
Incident wave & Plane wave & Plane wave \\
Electron energy & $300~\mathrm{keV}$ & $300~\mathrm{keV}$ \\ 
\hline
\end{tabular}
\label{tab:abtem_params}
\end{table} 

\section{Model Architecture Details}
\label{supp:training}
\paragraph{Computing Resources.} In our experiments, we leveraged high-performance computing (HPC) nodes equipped with 8 NVIDIA A100 GPUs with 80GB VRAM each and 128 CPU cores. 
\paragraph{VQGAN Training Details.}
We employ a 3D VQGAN as the first-stage autoencoder to learn a compact latent
representation for high-resolution scientific imaging volumes. The VQGAN is
trained on the complete multi-modal dataset, including all available imaging
domains and modalities, to learn a shared latent space across heterogeneous
scientific images. Importantly, the VQGAN training does not use any
modality-specific supervision or target-modality adaptation.

The input volume size is $16 \times 256 \times 256$. The encoder compresses each
volume into a latent representation with 8 channels and a spatial resolution of
$16 \times 32 \times 32$, which is subsequently used by the latent diffusion
model. The vector quantization module contains a codebook of 16,384 embeddings
with an embedding dimension of 8. The commitment cost is set to 0.25.

The VQGAN is trained for 1,000 epochs with a batch size of 1. We use Adam
optimization with learning rates of $3\times10^{-5}$ and $2\times10^{-5}$ for
the generator and discriminator, respectively. The training objective combines
reconstruction, perceptual, adversarial, and 3D feature-based Jukebox losses.
The perceptual, adversarial, and Jukebox loss weights are set to 0.1, 0.008,
and 0.7, respectively. Adversarial training is introduced after a 10-epoch
warm-up period. The discriminator is a 3D convolutional network with 64 channels and two
layers, using instance normalization. The VQGAN architecture uses three
encoder-decoder stages with channel dimensions of $[128,256,512]$ and two
residual layers per stage. All activation functions are ReLU.

\paragraph{Latent Diffusion Training Details.}
We train a latent diffusion model on the compact latent space learned by the
pretrained VQGAN. The diffusion model operates on latent representations with
8 channels and a volumetric resolution of $16 \times 32 \times 32$. All
experiments use precomputed latent representations extracted from the VQGAN
encoder rather than directly training the diffusion model in the image space.
The latent diffusion model is implemented using a 3D U-Net architecture with
channel dimensions of $[64,128,256,512]$ and two residual blocks at each
resolution level. Attention modules are applied at the two highest-resolution
stages. Cross-attention conditioning is enabled with a feature dimension of 64,
and the attention heads are configured with 32 channels at the corresponding
attention stages.

To enable scalable generation across heterogeneous imaging domains and
modalities, we adopt a universal compositional conditioning strategy. The
conditioning representation consists of three disentangled components:
domain, object, and contrast embeddings. Specifically, the model contains
3 domain categories, 12 object categories, and 9 contrast categories, with
each component represented in a 64-dimensional embedding space. The final
conditioning representation is obtained by combining these components,
allowing the model to generate unseen domain--object--contrast combinations
without additional target-modality training.
\begin{figure*}[htp]
    \centering
    \includegraphics[width=\textwidth]{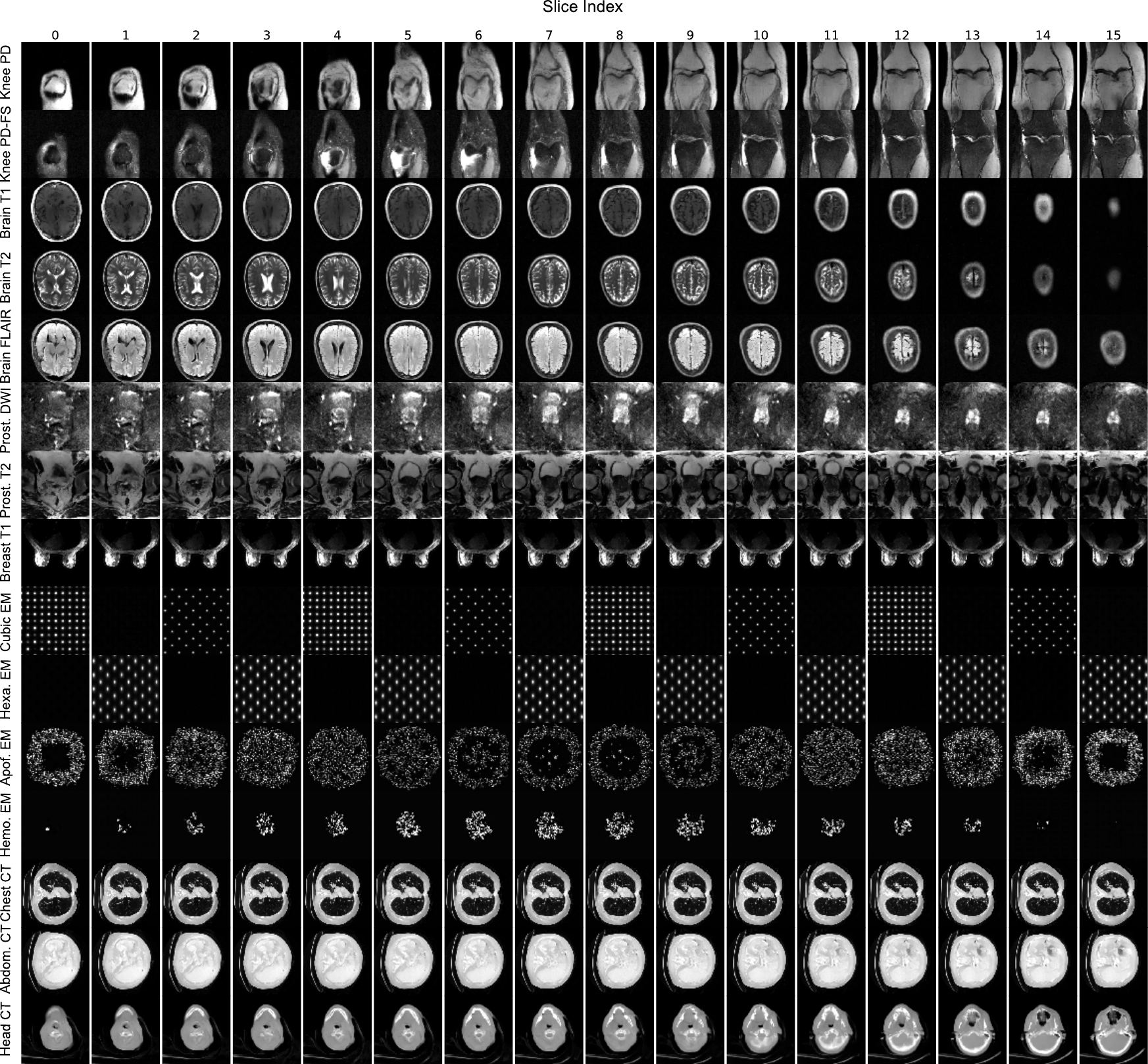}
    \caption{
   Representative generated 3D volumes across all imaging modalities.
Each row corresponds to one modality, and the 16 columns visualize axial
slices from a single generated volume.
    }
    \label{fig:generated_16frames}
\end{figure*}

The diffusion model is trained for 500 epochs with a batch size of 4 using
the Adam optimizer with a learning rate of $1\times10^{-4}$. The learning rate
is decayed with a factor of 0.9995 during training. We apply exponential
moving average (EMA) to stabilize optimization, with an EMA decay factor of
0.999.
We use a cosine noise schedule with 300 diffusion timesteps and adopt the
velocity ($v$)-prediction parameterization. The model is trained with a
unified objective across all imaging domains and modalities.
To incorporate domain-specific physical constraints, we introduce a
physics-guided regularization term during diffusion training. The physical loss
is activated using a warm-up strategy, starting from epoch 100 and gradually
increasing until epoch 150. The base physical loss scale is set to $5\times10^{-4}$.. This
warm-up strategy allows the diffusion model to first learn the general latent
distribution before introducing physics-based constraints.

\section{Additional Experiments}

While Table~\ref{tab:conditioning_comparison} in the main text reports results averaged across all
backbones for each domain, Table~\ref{tab:conditioning_domain_appendix}
provides the per-backbone breakdown underlying that average. Compositional
conditioning improves the majority of metrics in 10 of the 12
backbone--domain combinations with an available baseline, with the
largest gain observed for 3DMedDiff, which improves on all 15
metric--domain combinations. The exception is MedLord on MRI and EM,
where independently learned per-modality embeddings remain competitive,
suggesting that backbone-specific embedding capacity can partially
substitute for compositional structure in some settings. Fig.~\ref{fig:generated_16frames} visualizes generated volumes for all
modalities.
\begin{table*}[t]
\centering
\caption{
Domain-level comparison between backbone-specific baseline conditioning and
the proposed compositional conditioning across three imaging domains.
Results are macro-averaged over modalities within each domain, including eight
MRI modalities, four EM modalities, and three CT modalities.
For MedLord, the backbone-specific baseline uses an independently learned
embedding for each modality, whereas compositional conditioning uses the
proposed domain--object--contrast representation.
Other backbones retain their original conditioning mechanisms for the baseline
setting.}
\label{tab:conditioning_domain_appendix}

\setlength{\tabcolsep}{4.5pt}
\renewcommand{\arraystretch}{0.9}

\resizebox{\textwidth}{!}{%
\begin{tabular}{llccccc}
\toprule
\textbf{Backbone} &
\textbf{Domain} &
\textbf{VideoMAE-FD} $\downarrow$ &
\textbf{VideoMAE-KID} $\downarrow$ &
\textbf{I3D-FD} $\downarrow$ &
\textbf{I3D-KID} $\downarrow$ &
\textbf{JEDI} $\downarrow$ \\
&
&
&
$\times 10^3$
&
&
$\times 10^3$
& \\
\midrule

\multicolumn{7}{l}{\textbf{Backbone-specific baseline conditioning}} \\
\addlinespace[1pt]

\multirow{3}{*}{MedLord}
& MRI
& \underline{27.33}
& \underline{109.08}
& \underline{11.21}
& \underline{20.10}
& \underline{1.668} \\
& EM
& \underline{17.38}
& \underline{43.82}
& \underline{3.93}
& \underline{3.13}
& \underline{2.522} \\
& CT
& 47.13
& 184.68
& 23.85
& 47.63
& 3.969 \\
\cmidrule(lr){1-7}

\multirow{3}{*}{Pixel Diffusion}
& MRI
& 45.84
& 141.80
& \underline{29.28}
& \underline{63.99}
& 2.042 \\
& EM
& 87.48
& 297.31
& \underline{9.81}
& \underline{15.93}
& 10.519 \\
& CT
& 45.48
& 150.33
& 28.85
& 54.51
& 3.66 \\
\cmidrule(lr){1-7}

\multirow{3}{*}{3DMedDiff}
& MRI
& 107.39
& 459.01
& 42.07
& 102.67
& 8.092 \\
& EM
& 35.74
& 110.34
& 5.94
& 7.27
& 5.480 \\
& CT
& 122.30
& 497.63
& 46.92
& 108.90
& 12.816 \\
\cmidrule(lr){1-7}

\multirow{3}{*}{Medical Diffusion}
& MRI
& 36.97
& 150.29
& 14.30
& \underline{27.09}
& 1.694 \\
& EM
& 22.64
& 59.03
& 5.34
& 7.68
& 3.769 \\
& CT
& 43.99
& 171.12
& \underline{20.68}
& 37.90
& \underline{4.120} \\

\midrule

\multicolumn{7}{l}{\textbf{Compositional conditioning}} \\
\addlinespace[1pt]

\multirow{3}{*}{MedLord}
& MRI
& 27.45
& 109.88
& 11.65
& 21.76
& 1.718 \\
& EM
& 18.49
& 51.36
& 4.14
& 4.09
& 2.662 \\
& CT
& \underline{38.67}
& \underline{153.33}
& \underline{18.88}
& \underline{35.00}
& \underline{3.642} \\
\cmidrule(lr){1-7}

\multirow{3}{*}{Pixel Diffusion}
& MRI
& \underline{40.41}
& \underline{125.66}
& 29.36
& 64.10
& \underline{1.949} \\
& EM
& \underline{84.43}
& \underline{282.65}
& 9.89
& 16.43
& \underline{9.889} \\
& CT
& \underline{39.07}
& \underline{128.17}
& \underline{27.56}
& \underline{52.13}
& \underline{3.64} \\
\cmidrule(lr){1-7}

\multirow{3}{*}{3DMedDiff}
& MRI
& \underline{27.97}
& \underline{110.11}
& \underline{12.17}
& \underline{22.46}
& \underline{1.379} \\
& EM
& \underline{22.49}
& \underline{64.49}
& \underline{4.06}
& \underline{3.73}
& \underline{3.593} \\
& CT
& \underline{29.65}
& \underline{107.37}
& \underline{17.16}
& \underline{28.55}
& \underline{3.530} \\
\cmidrule(lr){1-7}

\multirow{3}{*}{Medical Diffusion}
& MRI
& \underline{35.86}
& \underline{146.07}
& \underline{14.26}
& 27.16
& \underline{1.690} \\
& EM
& \underline{22.50}
& \underline{58.82}
& \underline{4.97}
& \underline{6.32}
& \underline{3.644} \\
& CT
& \underline{43.35}
& \underline{166.60}
& 20.81
& \underline{37.55}
& 4.299 \\
\cmidrule(lr){1-7}

\multirow{3}{*}{SciFusion (Ours)}
& MRI
& \textbf{22.95}
& \textbf{87.56}
& \textbf{9.99}
& \textbf{17.11}
& \textbf{1.43} \\
& EM
& \textbf{18.40}
& \textbf{50.11}
& \textbf{4.63}
& \textbf{5.19}
& \textbf{1.57} \\
& CT
& \textbf{31.81}
& \textbf{123.55}
& \textbf{16.72}
& \textbf{30.38}
& \textbf{3.08} \\

\bottomrule
\end{tabular}%
}

\vspace{2pt}
\begin{minipage}{0.98\textwidth}
\footnotesize
\textit{Note:}
Underline values indicate the better result between baseline and compositional
conditioning for the same backbone and domain.
\end{minipage}
\end{table*}

\end{document}